\documentclass{aa}
\usepackage{graphicx}
\usepackage{epstopdf}
\usepackage{amssymb}
\DeclareGraphicsExtensions{.ps,.eps,.jpg,.pdf}
\usepackage{epsfig,times}
\usepackage{color}
\usepackage{txfonts,ulem}

\begin{document}
 
\title{r-Java : An r-process Code and Graphical User Interface for Heavy-Element Nucleosynthesis
}
 
\author{Camille Charignon~\inst{1,2}, Mathew Kostka~\inst{1}, Nico Koning~\inst{1}, Prashanth Jaikumar~\inst{3,4} and Rachid Ouyed~\inst{1}\thanks{Email:rouyed@ucalgary.ca}}

\institute{
$^1$Department of Physics \& Astronomy, University of Calgary, 2500 University Drive NW, Calgary, AB T2N 1N4, Canada\\
$^2$Department de Physique, Ecole Normale Sup\'{e}rieure de Cachan, Cachan 94230, France\\
$^3$Department of Physics \& Astronomy, California State University Long Beach, 1250 Bellflower Blvd., Long Beach, California 90840 USA\\
$^4$Institute of Mathematical Sciences, CIT Campus, Chennai 600113, India}

\date{Received <date>; accepted <date>}

\authorrunning{Charignon et al.}

\titlerunning{r-process Nucleosynthesis with r-Java} 

\abstract{We present r-Java, an r-process code for open use, that performs r-process nucleosynthesis calculations.  Equipped with a simple graphical user interface, r-Java is capable of carrying out nuclear statistical equilibrium (NSE) as well as static and dynamic  r-process calculations for a wide range of input parameters. In this introductory paper, we present the motivation and details behind r-Java, and results from our static and dynamic simulations. Static simulations are explored for a range of neutron irradiation and temperatures. Dynamic simulations are studied with a parameterized expansion formula. Our code generates the resulting abundance pattern based on a general entropy expression that can be applied to degenerate as well as non-degenerate matter, allowing us to track the rapid density and temperature evolution of the ejecta during the initial stages of ejecta expansion. At present, our calculations are limited to the waiting-point approximation. We encourage the nuclear astrophysics community to provide feedback on the code and related documentation, which is available for download from the website of the Quark-Nova Project: http://quarknova.ucalgary.ca/ }
\maketitle



\section{Introduction}

The rapid  neutron capture process  (r-process) is believed to  be the
mechanism for the nucleosynthesis of about half of the stable nuclei
heavier than  Iron (\cite{Burbidge,Cameron}). Explosive and neutron-rich
astrophysical environments present  ideal conditions for the r-process
to take  place. Possible candidate  sites discussed in  the literature
include  the  neutrino-driven neutron-rich wind from proto-neutron
stars (\cite{Qian96,Qian03b}), prompt explosions of collapsed stellar cores, (\cite{Sumiyoshi,Wanajo03}),  neutron  star  decompression (\cite{Meyer,Goriely}), tidal disruption in binary merger events (\cite{FRT}), outflows in Gamma-Ray Bursts (\cite{Surman}) etc.  Most  importantly,  abundance data on
r-process elements in  metal-poor stars (\cite{Sneden03}) and certain radionuclides in meteorites (\cite{QW08}) point   toward  the  distinct   
possibility  of multiple r-process sites, so that more than
one of these sites may contribute to the observed abundance 
pattern of the r-process elements (\cite{Truran}).  

\vskip 0.2cm

In 1985, an important work by Bethe \& Wilson  
brought neutrino-driven winds from Type II Supernovae (SNe) to the forefront 
of the discussion about astrophysical sites for the r-process. Since then, much progress has been made in the modelling of type II SNe and neutrino winds of nascent neutron stars. Some examples of such work
are Woosley et al. (1994), Takahashi et al. (1994), Qian \& Woosley (1996), Cardall \& Fuller (1997),
Otsuki et al.  (2000), Wanajo et al. (2001), Thomson et al. (2001). 
However, we have not so far arrived upon a single self-consistent
astrophysical model that reproduces the observed abundance of
r-process elements and at the same time, is consistent with spectroscopic and meteoritic data. The most natural explanation is that the observed r-process pattern follows from a superposition of 
neutron capture events with differing neutron-to-seed ratios and
exposure time-scales. Producing the third peak requires
extreme values of entropy and dynamic time-scale that are a challenge for current models of Type II SNe explosions~(\cite{QW08}). Neutron star mergers provide a much larger neutron-to-seed ratio than type II SNe,
but occur so infrequently that they fail to explain the early enrichment 
of r-process elements relative to Iron observed in metal-poor stars
(\cite{Qian00}).  Clearly, further study is needed to better understand the r-process conditions, 
to this end we present a fairly general and flexible r-process code that is transparent and freely available to the nuclear physics and astrophysics community.

\vskip 0.2cm

\begin{figure*}[t!]
\centering
\includegraphics[width=0.65\textwidth]{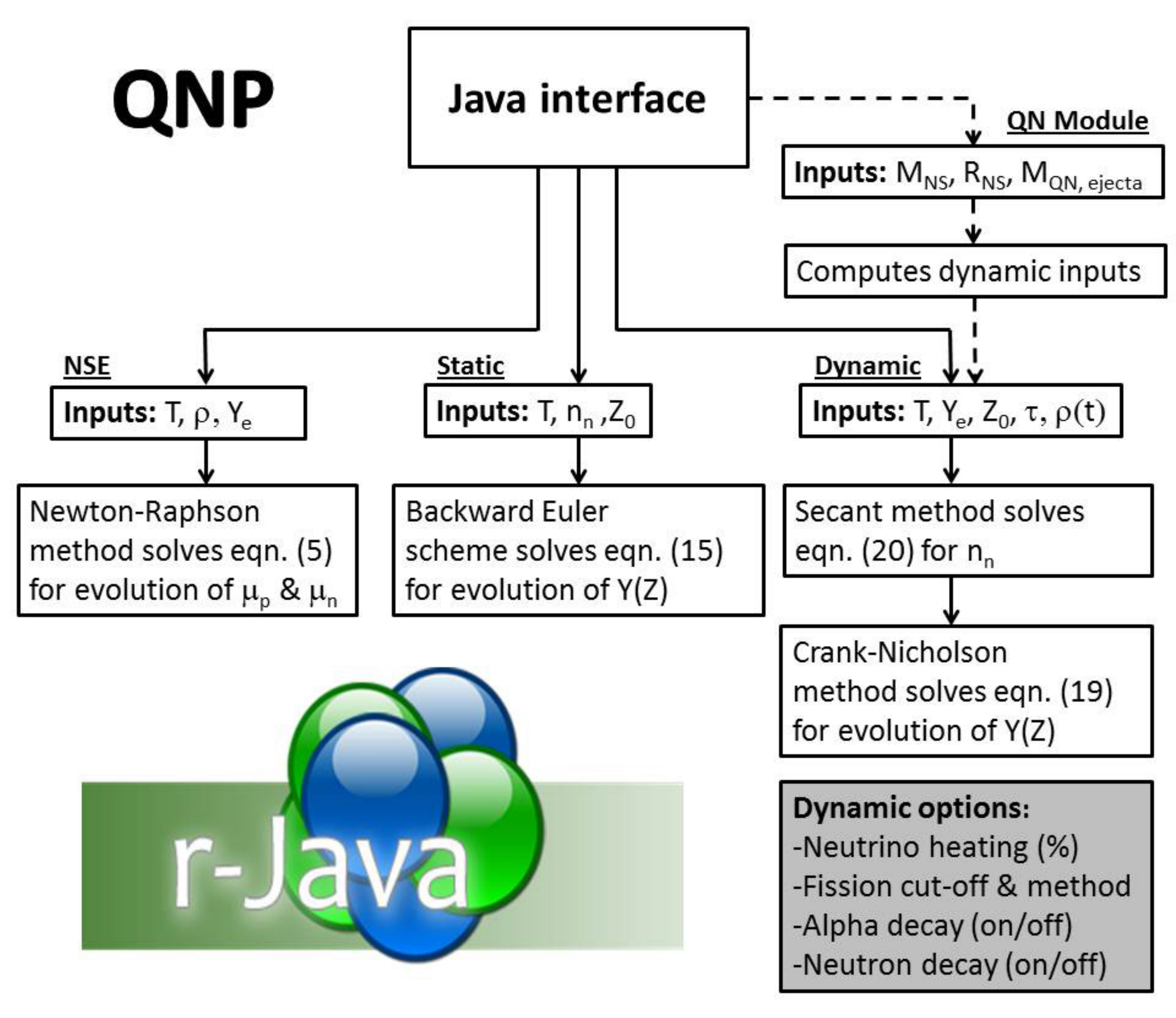}
\caption{Schematic representation of r-Java's capabilities and tools (see text for meaning of symbols). More details can be found in the manual available on the quark nova project website: http://quarknova.ucalgary.ca/}
\label{rjavachart}
\end{figure*}

 The motivation behind developing our own r-process code came in part from recognizing the difficulty entailed in having access to an openly shared cross-platform program where any interested researcher can examine and modify the dynamical evolution equations and the nuclear input.  We perceive a distinct need for this, as rare-isotope experiments such as FRIB later in this decade will make available new and improved nuclear data for reaction rates relevant to the r-process. Besides, the r-process is of wide interest to the astrophysics community who might wish to explore different dynamical environments for the r-process, and not be as concerned, so to speak, about the input nuclear physics. To avoid a kind of "black-box" situation arising from limited exchange of numerical codes, we decided to build a new code that is flexible enough to incorporate different conditions and has detailed documentation. The code is capable of simulating nuclear statistical equilibrium (NSE), neutron bombardment of a static target as well as the expansion (including the relativistic case) of neutron-rich material. In this work, we discuss all of the above and also apply this code to the unique environment of an ejected and expanding neutron star crust. For the code we designed a graphical user interface (GUI) which allows for real-time interpretation and analysis of simulations through data and graphs.  We have made the r-process application freely available in Java-based format on the quark nova project website\footnote{ http://quarknova.ucalgary.ca/software/rJava/ . If using this code for calculations for a paper, we request that you please acknowledge this website and cite the online arXiv article, or the published version when it becomes available.}, where one can also find input nuclear data and supporting documentation on most aspects of our code.  We request users to test it, and welcome questions and feedback. Our aim is to make our nucleosynthesis code, which we call r-Java, and its workings as transparent as possible so that any scientific user can adapt it to their needs.

\vskip 0.2cm

This paper is organized as follows.  In section 2, we mention our sources for the nuclear input within r-Java and display results for NSE that are in agreement with previous benchmarks. In sections 3 and 4, we present the simplified reaction network (waiting-point approximation) and simulation results for heavy-element production by static and dynamic neutron capture. In section 5, the conclusions from our simulations are discussed and we look ahead to future work.

\section{The R-process Code: r-Java}

\begin{figure}[b!]
    \includegraphics[width=0.5\textwidth]{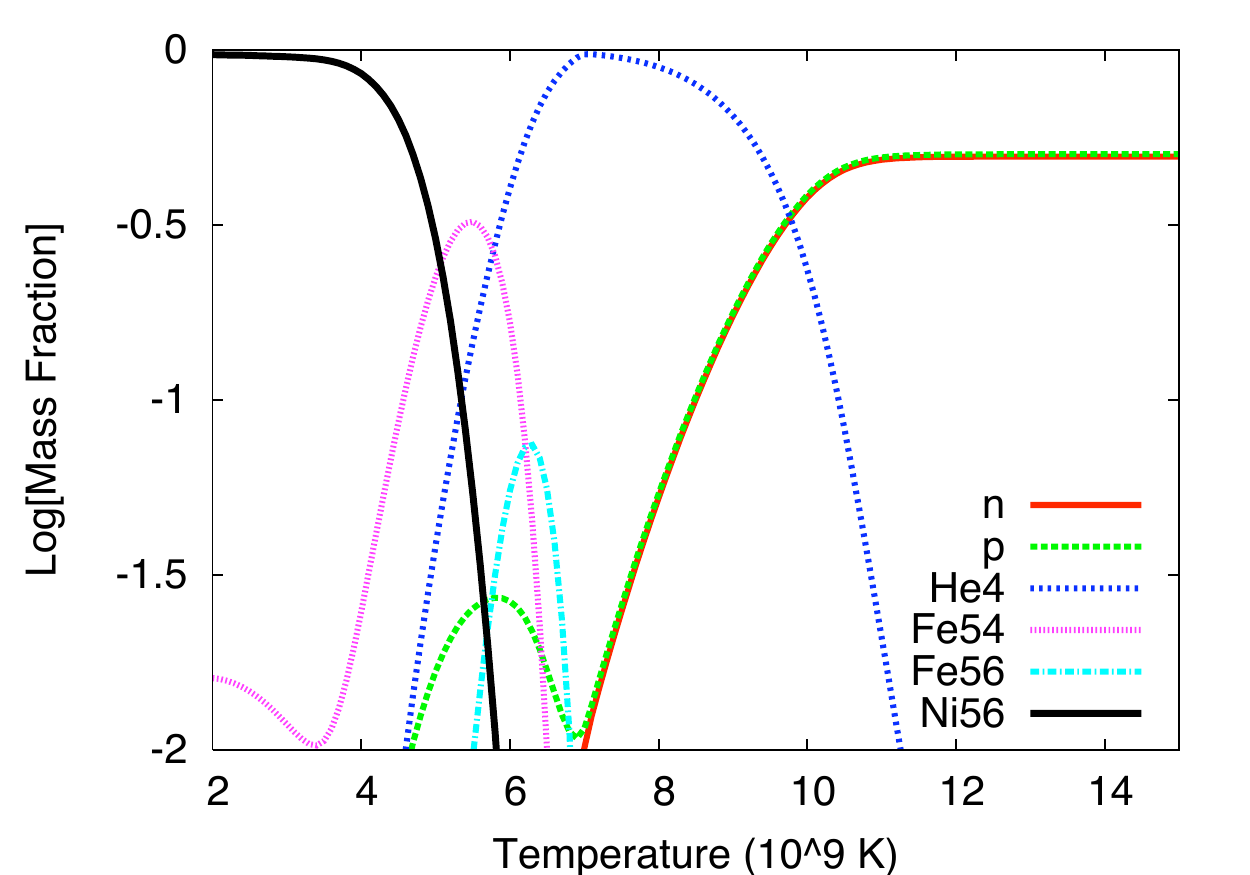}
    \caption{Abundances of nucleons and selected nuclei in the NSE state with varying temperature, at $\rho=1\times 10^{7} g.cm^{-3}$ and $Y_{e}=0.5$}
    \label{NSEvsT}
 
   \includegraphics[width=0.5\textwidth]{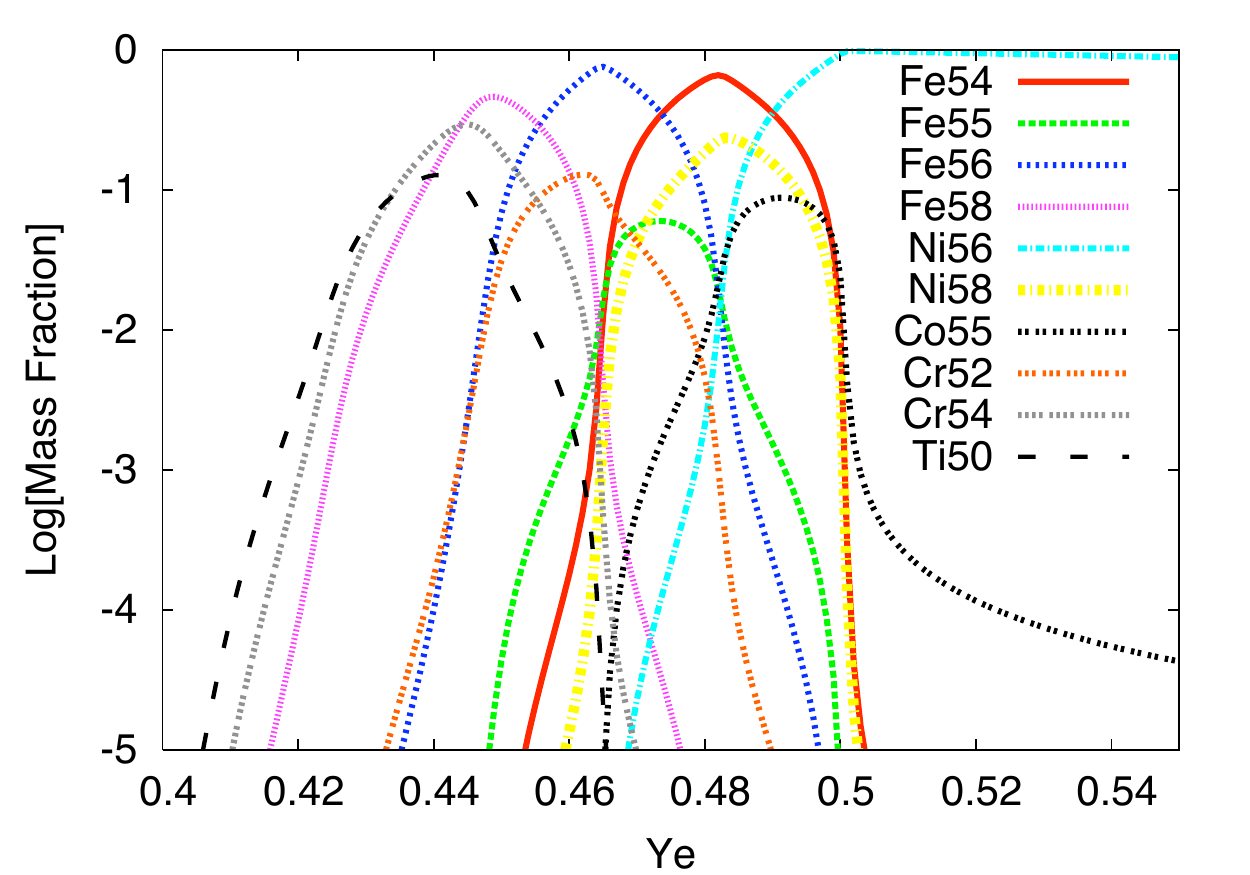}
    \caption{Abundances of selected nuclei in the NSE state with a varying electron-to-baryon fraction, at $\rho=1\times 10^{7} g.cm^{-3}$ and $T=3.5 \; T_{9}$}
    \label{NSEvsYe}

    \includegraphics[width=0.5\textwidth]{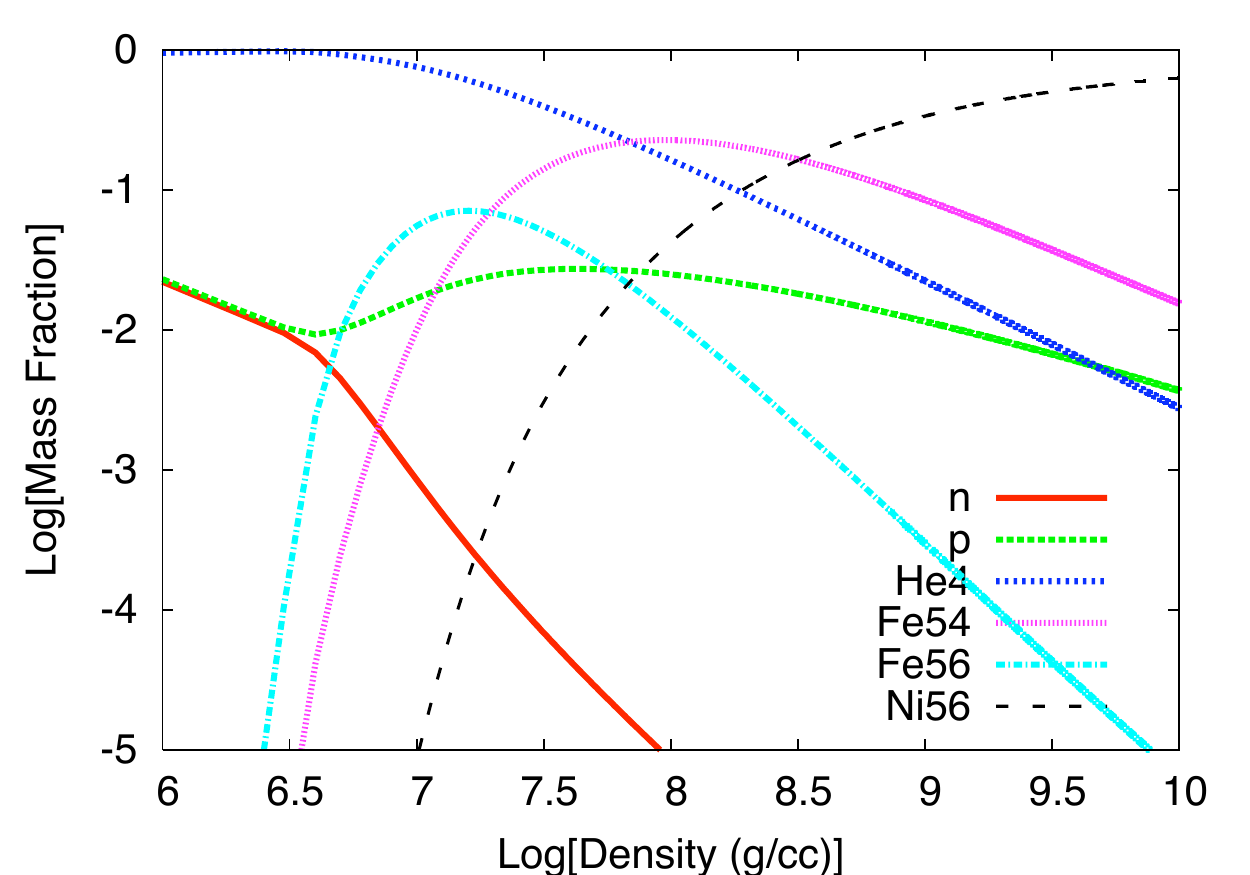}
    \caption{Abundances of nucleons and selected nuclei in the NSE state with varying density, at $Y_e=0.5$ and $T=6.5 \; T_{9}$}
    \label{NSEvsRho}
 \end{figure}

Fundamental to our code is the waiting point approximation which sets lower limits on the temperature of $2\times10^9$K and neutron density of $10^{20}$ cm$^{-3}$ for all our simulations (\cite{Cameron83}). Each of the modules of r-Java; namely NSE, static and dynamic have  their own appropriate set of input parameters and numerical solution methods (see Fig. \ref{rjavachart}).  For example, in the dynamic case, the user can choose the amount of heating from neutrinos, turn on/off nuclear fission, alpha or neutron decay and specify the density profile of the expanding material.  The nuclear input data that is provided along with r-Java uses nuclear masses from experimental sources as much as possible (\cite{ExpMasses}), and where none was available we used \cite{Bdecay}.  For the two nuclear inputs: the neutron separation energy ($S_{n}$) and the $\beta$-decay rates ($\lambda^{\beta}$), we used \cite{Bdecay}.  A freedom provided to the users of r-Java is the ability to modify any nuclear input in order to simulate the effect that such a change would have on abundance yields.  There are many different outputs that can be generated by r-Java as well as the ability to animate how a given parameter or abundance changes during a simulation run.  A unique module included in r-Java displays the periodic table and in real-time shows which elements are being created and destroyed during a simulation.  Also included is a module to study nucleosynthesis in a Quark-Nova~(\cite{Jai07}), which will be discussed in detail in subsequent papers. The following subsections are dedicated to describing in detail the physics involved in each of the modules of r-Java and presenting corresponding simulation results.

\subsection{Nuclear Statistical Equilibrium}

As a prelude to more complex situations, we begin by developing a NSE code where the determining nuclear property is the binding energy. Consider an astrophysical plasma composed of photons, free neutrons and protons, and a mix of seed nuclei such as $^{56}$Fe, with temperature and density high enough that the nuclear reactions assembling nuclei from free neutrons and protons are much faster than the expansion time scale. For large enough temperatures $T_9$=$T/(10^9)$K$\simeq 6$, the system would then reach NSE with forward and backward reaction rates equal. As long as such conditions prevail, the detailed balance equation holds 

\begin{equation}
\label{detailbalance}
Z_{i}\mu_{p}+N_{i}\mu_{n}=\mu_{i} \ .
\end{equation}

where $\mu_i$ is the chemical potential of the species $i$ (with $Z_{i}, N_{i}, A_{i}$
 the corresponding atomic number, neutron number, and mass number, respectively). Assuming all the nuclei in the gas follow Maxwell-Boltzmann statistics, the particle number densities are

\begin{equation}
\label{numberdensity}
n_{i}=g_{i} \left(\frac{2\pi kT m_{i}}{h^2}\right)^{3/2} \exp{\left(\frac{\mu_{i}+B_{i}}{kT} \right)} \ .
\end{equation}

where $B$ is the binding energy and $g$ is the degeneracy factor. In addition, mass and charge conservation imply that

\begin{equation}
\label{massconservation}
\sum_{i} A_{i} Y_{i} = 1 \ ,
\end{equation}

\begin{equation}
\label{chargeconservation}
\sum_{i} Z_{i} Y_{i} = Y_{e} \ .
\end{equation}

Combining Eqs. (\ref{numberdensity}) and (\ref{detailbalance}) with the conservation equations (\ref{massconservation}) and (\ref{chargeconservation}), we obtain the following equations for the neutron and proton chemical potentials:

\begin{eqnarray}
\left\{ \begin{array}{l}
 \sum_{i} A_{i} n_{i}( \mu_{p},\mu_{n}) - \rho N_{A} = 0  \\
 \sum_{i} Z_{i} n_{i}( \mu_{p},\mu_{n}) - Y_{e} \rho N_{A} = 0  
\end{array} \right.
\end{eqnarray}

For any given $Y_e$ (electron fraction), $T$ (the temperature in Kelvin) and baryon mass density $\rho$ (in g cm$^{-3}$), we then solve this system by the Newton-Raphson method for $ \mu_{n} $ and $\mu_{p}$ (Fig. \ref{rjavachart}), which we use to obtain the mass fractions of every nucleus through Eqn.(\ref{massabundance}), where $N_A$ is the Avogadro number. The nuclei that dominate the abundances distribution in NSE depend only on $\rho,T$ and $B$. High densities favor large nuclei due to the $\rho$-dependence while high temperatures favor light nuclei since the Planck distribution contains energetic photons which photo-disintegrate heavy nuclei. We found the detailed balance method (Eqn.~(\ref{detailbalance})) numerically amenable to the Newton-Raphson method. However, we should caution that the success of the Newton-Raphson method depends sensitively on a proper choice for the initial value of  $\mu_p,\mu_n$ and the starting composition.  We chose to use the NSE calculations of \cite{NSEcode} as a test of our code. We have plotted our NSE results in Fig.\ref{NSEvsT}. We also display the mass fractions as a function of $Y_e$ in Fig.\ref{NSEvsYe}. Comparing the former to the results of \cite{CliffTayler}, and the latter with Figure 3 of \cite{NSEcode}, we find very good agreement with these established results. In Fig.\ref{NSEvsRho}, we have also shown a new plot: the mass fractions obtained with our code as a function of density, for fixed $Y_e=0.5$ and $T=6.5\times 10^9$K.

\section{Reaction Network and Static Case}

\begin{figure*}[t!]
\includegraphics[width=180mm]{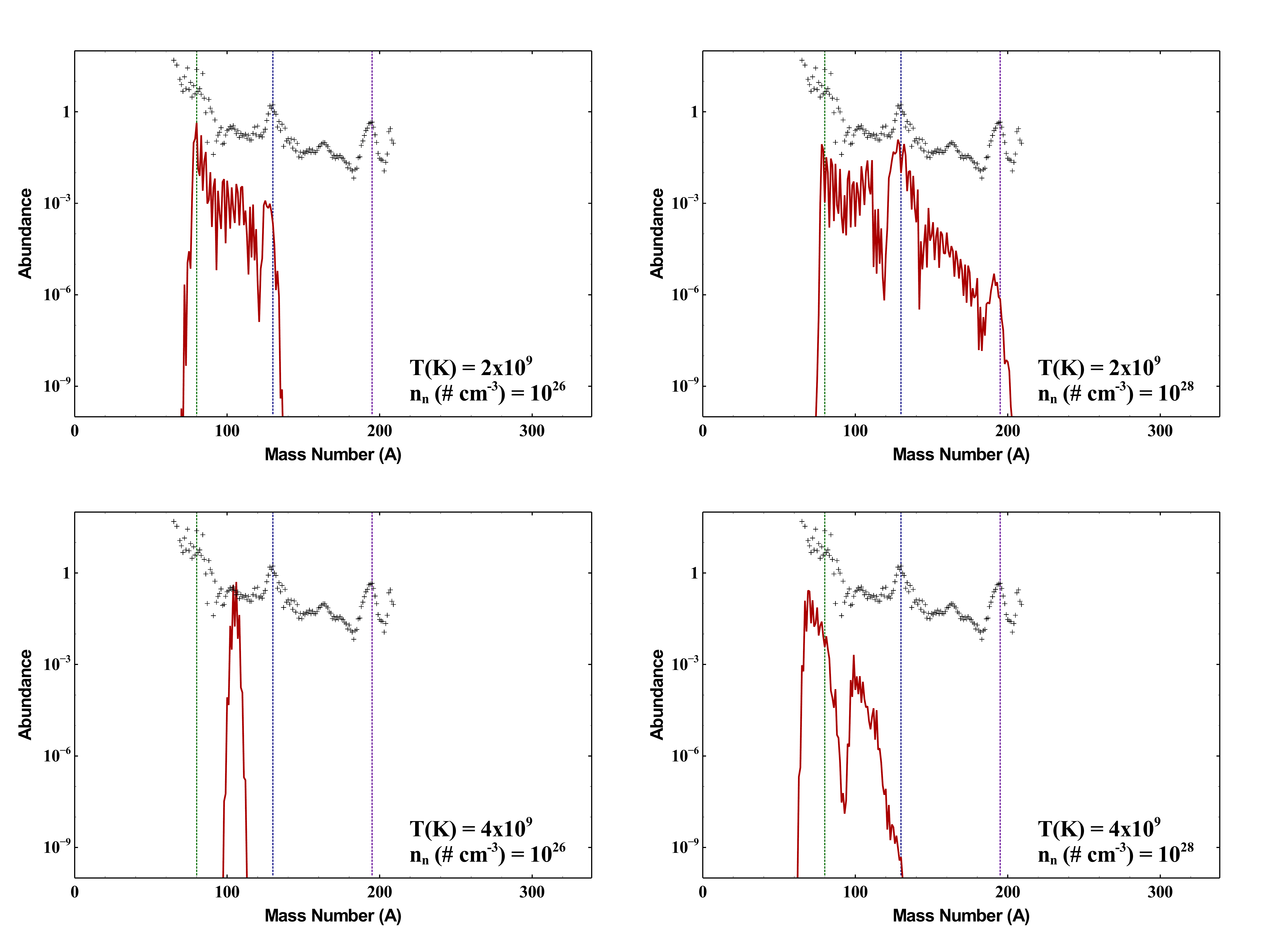}
\caption{Final abundances in the static simulations after one second of neutron bombardment and at different physical conditions. {\bf Top left:}  $n_{n}=10^{26}$cm$^{-3}$ and $T_9$=2, {\bf Top right:}  $n_{n}=10^{28}$cm$^{-3}$ and $T_9$=2, {\bf Bottom left:}  $n_{n}=10^{26}$cm$^{-3}$ and $T_9$=4, {\bf Bottom right:}  $n_{n}=10^{28}$cm$^{-3}$ and $T_9$=4. The solar abundance is shown in each case with vertical lines marking the location of the important peaks.}
\label{StaticResults}
\end{figure*}

\begin{figure*}[t!]
\includegraphics[width=180mm]{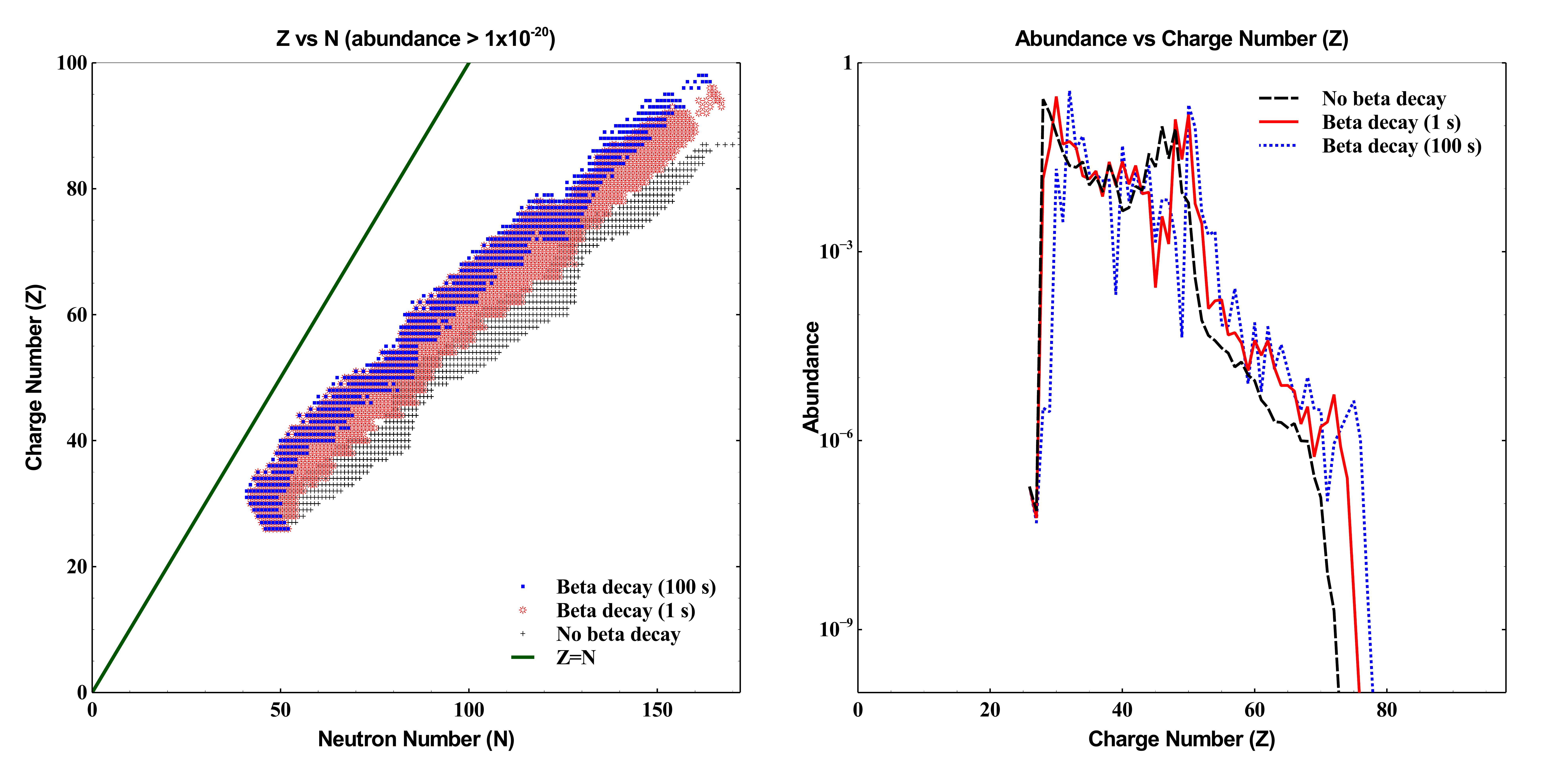}
\caption{Final abundances in the static simulations showing how the elements $\beta$-decay back to stability once bombardment of neutrons stops. The cases shown are one second and one hundred seconds following the end of neutron bombardment.  The left panel is the corresponding  Z versus N distribution while the right panel displays how the abundances versus charge number evolve in time.}
\label{StaticResultsBetaDecay}
\end{figure*}

In general, a whole variety of reactions can occur simultaneously, producing and destroying nucleus $i$, whose reaction induced density evolution is described by:
\begin{equation}
\label{reacnet}
\left(\frac{dn_{i}}{dt}\right)_{\rho = {\rm const}} = \sum_{j} N_{j}^{i} r_{j} + \sum_{j,k} N_{j,k}^{i} r_{j,k} \ ,
\end{equation}

where one-body reactions, such as decays and photo-disintegrations (eg. $ r_{j}=\lambda_{j} n_{j}$) and two body reactions ($r_{jk}= \frac{1}{1+\delta_{jk}} <\sigma v>_{jk} n_{j} n_{k}$ where $\delta_{jk}$ is
 the Kronecker delta) dominate. The individual $N^{i}s$ in Eqn.~(\ref{reacnet}) are positive or negative numbers specifying creation and annihilation. Then, the number abundances $Y_{i}$ and mass abundances $X_i$ are given by
\begin{eqnarray}
Y_{i} &=& \frac{n_{i}}{\rho N_{A}} \,,\\
X_{i} &=& \frac{m_{i}n_{i}}{\rho N_{A}} \,.
\label{massabundance}
\end{eqnarray}

In terms of $Y_{i}$, the nuclear reaction network equation (\ref{reacnet}) becomes

\begin{equation}
\label{abunnet}
\dot{Y}_{i} = \sum_{j} N_{j}^{i} \lambda_{j} Y_{j} + \sum_{j,k} \frac{N_{j,k}^{i}}{1+\delta_{jk}} \rho N_{A} <\sigma v>_{jk} Y_{j} Y_{k} \ .
\end{equation}

Applying this reaction network to the r-process, we will limit ourselves at the moment to the following interactions:

\begin{itemize}
\item neutron capture
\item photo-disintegration
\item $\beta$-decay
\end{itemize}

For the static simulation, we assume a static, non-expanding chunk of neutron-rich material near drip density ($\approx 4\times 10^{11}$g/cc), described by the BPS equation of state\footnote{This leaves only T, n$_n$ and Z$_0$ (the dominant nucleus at that density) as inputs since Y$_e$ is prescribed.} (\cite{Crust}), being bombarded by a large and constant flux of neutrons over a given period of time. Eqn. (\ref{abunnet}) is then

\begin{eqnarray}
\label{reducenet}
\dot{Y}(Z,A) &=& \lambda_{Z-1,A}^{\beta} Y(Z-1,A) \;+\; <\sigma v> n_{n} Y(Z,A-1)  \nonumber \\
&+& Y(Z,A+1)\lambda_{Z,A+1}^{\gamma} - Y(Z,A)(<\sigma v> n_{n} \nonumber \\
 &+& \lambda_{Z,A}^{\gamma} \;+\; \lambda_{Z,A}^{\beta}\:)  \ .
\end{eqnarray}

\noindent where $\langle\sigma v\rangle$ is the thermally averaged neutron capture cross-section. Since we have a high neutron density  ($10^{22}$-$10^{26}$cm$^{-3}$) and high temperatures ($T\gtrsim 2\times 10^{9} K$), neutron captures and photo-disintegrations occur much faster than $\beta$-decays. For such conditions we can assume that $(n,\gamma)\rightleftharpoons (\gamma,n)$ equilibrium is established within every isotopic chain. We also assume steady flow while the high density and temperature conditions remain, viz.,
\begin{equation}
\label{ngEqu}
\dot{Y}(Z,A)=0 \ .
\end{equation} 
\noindent This leads to steady $\beta$-flow or the standard waiting-point approximation in which we can express the abundance ratio of two neighboring isotopes within one isotopic chain in terms of only the neutron separation energy. In terms of the reduced reaction network (\ref{reducenet}), we then have:

\begin{eqnarray}
\label{abunratio}
\lefteqn{\frac{Y(Z,A+1)}{Y(Z,A)} = } \nonumber \\
&& n_{n}\frac{g(Z,A+1)}{2g(Z,A)}\left(\frac{A+1}{A}\right)^{3/2}\left(\frac{2 \pi \hbar^2}{\mu kT}\right)^{3/2} e^{\frac{S_{n}(Z,A+1)}{kT}} \ .
\end{eqnarray}

\noindent Further, we introduce the abundance of an isotopic chain, $Y(Z)$, defined as the sum of every isotope abundance in that chain: $ Y(Z) = \sum_{A} Y(Z,A)$. With Eqn.(\ref{abunratio}), we can now express every $Y(Z,A)$ in terms of only the corresponding $Y(Z)$ and population coefficients defined through $Y(Z,A)=P(Z,A)Y(Z)$ which depend only on $n_{n}$, T and $S_{n}$. This allows us to dramatically reduce the number of equations from several thousand down to 136, the number of isotopic chains we have in our network. Now, we study the evolution of Y(Z) through $\beta$-decay:

\begin{eqnarray}
\label{eq:Yevolv1}
\dot{Y}(Z) &=& \sum \dot{Y}(Z,A) \nonumber \\
           &=& \sum(\lambda_{Z-1,A}^{\beta} Y(Z-1,A) - \lambda_{Z,A}^{\beta} Y(Z,A) )\ .
 \end{eqnarray}
 This leads to
 \begin{eqnarray}
   \dot{Y}(Z)  &=& \sum(\lambda_{Z-1,A}^{\beta} P(Z-1,A)) \times Y(Z-1) \nonumber \\
           &&  - \sum(\lambda_{Z,A}^{\beta} P(Z,A)) \times Y(Z)\ . 
\end{eqnarray}

\noindent Introducing effective $\beta$-decay rates: $\lambda_{Z}^{eff}=\sum(\lambda_{Z,A}^{\beta} P(Z,A))$, we obtain a system of coupled differential equations:

\begin{equation}
\label{EvoEqu}
\dot{Y}(Z) = \lambda_{Z-1}^{eff} Y(Z-1) - \lambda_{Z}^{eff} Y(Z) \ .
\end{equation}

\noindent As the abundances $Y(Z)$ can vary by many orders of magnitude within the integration interval, equation (\ref{EvoEqu}) is a very stiff system, so we use an implicit scheme, the Backward Euler Scheme, to treat the problem.
 To present the capabilities of r-Java in the static regime we varied the neutron density number and the temperature of the material to observe their impact on the r-process abundance (see Fig. \ref{StaticResults}).

Comparing the left and right figures in the top panel of Fig.~\ref{StaticResults} shows clearly that the third r-process peak at A=195 is realized when the neutron number density is high ($n_{n}=10^{28}$cm$^{-3}$) and the temperature is close to $2\times10^9$K. In reality, in dense neutron-rich environments neutron number densities above $10^{32}$cm$^{-3}$ are possible, and we might expect large $A$ nuclei to be copiously produced, but as we have not included fission in the static simulations, we have restrained our code at present to lower neutron densities ($n_{n}=10^{26}-10^{28}$cm$^{-3}$), so that nuclei do not build up without limit.

\subsection{Static Simulation Results}

From the top panels of Fig.~\ref{StaticResults}, it can be seen that our static simulations produce a weak third r-process peak at A=195 even when the neutron number density is relatively high ($n_{n}=10^{28}$cm$^{-3}$). Higher neutron densities can produce a strong third peak ($n_{n}\sim10^{32}{\rm cm}^{-3}$), but the absence of fission presently limits our static code to moderate neutron densities. Absence of fission cycling also leads to a broad peak structure. The bottom panels of Fig. ~\ref{StaticResults} show the effects of doubling the temperature, relative to the top panel, keeping neutron density the same. As temperature increases heavy nuclei tend to disappear since photo-disintegrations of neutron-rich nuclei become highly effective freezing the nuclear flow at lower A. In this case, the peaks in the mass distribution appear at lower mass numbers compared to the corresponding usual r-process peaks. This was found to be the case even for longer periods of neutron exposure, possibly suggesting that photo-disintegrations are dominating over neutron-captures within an isotopic chain, violating the conditions of $(n,\gamma)$ equilibrium. Such a "shift" of the peaks to incorrect locations in mass number has been noticed in earlier dynamical situations, such as neutron star mergers (eg. \cite{Frei}), where it is attributed to the choice of $Y_e$. In our static simulations, $Y_e$ does not cause the shift. Rather, in our case, this effect is driven by the high temperature, which favours photo-disintegrations and takes the r-process path closer to the valley of beta-stability, thereby downplaying the role of waiting-point nuclei in determining the abundance peaks (see \cite{goriely96} for the high neutron density and low temperature, $T_{9} <2$, regime). Extremely high neutron densities are required to restore the peak to its correct position. 

\subsection{The $\beta$-decay module}

Although we have implemented the waiting-point approximation, r-Java also provides the option of evolving the elements by $\beta$-decay back to stability from the time bombardment of neutrons is abruptly terminated. Thus, one can isolate key $\beta$-decay rates for  particular reaction networks and studies of freeze-out, which is known to be important, for eg., in explaining the rare-earth peak near Europium~(\cite{Sur97}). We use the Euler implicit scheme for each isotopic chain of elements with decay rate of each element exactly  the ones we used to compute the effective rates defined in Eqn.(\ref{EvoEqu}). 
Shown in Figure \ref{StaticResultsBetaDecay} are final abundances from the static simulation (under same
input conditions as in Fig. \ref{StaticResults}) $\beta$-decaying one second and one hundred seconds after the bombardment of neutrons has stopped. An added convenience in r-Java is the fact that the $\beta$-decay module can be run independent of nucleosynthesis calculations by specifying the initial composition as a vector.  We added a modification of the bi-diagonal solver, so that this solver accepts a vector of variable size. Currently the $\beta$-decay module  is limited to nuclei with mass between 50 and 300.

\section{Dynamic Case}
\label{DynamicSims}

The dynamic module of r-Java allows the user to input the initial temperature ($T_0$), density ($\rho_0$), electron fraction ($Y_{\rm e,0}$) and dominantly abundant element (Z$_0$) of a chunk of material.  The material then expands at a user specified expansion time-scale ($\tau$) for a given duration.  In order for r-Java to accommodate a wide variety of ejection mechanisms we have allowed the density profile ($\rho\left(t\right)$) to be freely modified by the user\footnote{The default density profile in the dynamic module of r-Java comes from studies of non-relativistic decompression by \cite{DensityEvo}:

\begin{equation}
\label{density}
  \rho (t) = \frac{\rho_{0}}{(1+\frac{t}{2 \tau})^2} \ 
\end{equation}}.

 The relativistic expansion regime will be discussed in an upcoming paper, here we discuss the non-relativistic case.  Users of r-Java are given the choice to include a simple fission model for dynamic simulations, in which the last species of the network can be specified.  For our dynamic simulations we assume that the last species in the network is $Z$=$92$, $A$=$276$  which fissions into two fragments--one with $Z$=$40$ and one with $Z$=$52$. The code does not presently include other types of fission (neutron-induced, neutrino-induced, $\beta$-delayed) or $\beta$-delayed neutron emission. Further options given to the user of r-Java for dynamic simulations are the ability to specify the fraction deposited from $\beta$-decays, and allowing $\beta$-decay and neutron decay to be turned on or off.

 Once the user provides the input parameters as above,
 the initial  entropy/nucleon is determined (in $k_B$=1 units) from the initial density, electron fraction and temperature using

\begin{equation}
\label{Entropy}
s = \frac{11 \pi^2}{45}\frac{T^3}{\rho}+ \frac{\mu_e^2T}{3\rho}+\frac{5}{2m_N}+\frac{1}{m_N}{\rm ln}\left(\frac{g_0(2\pi m_Nk_BT)^{3/2}}{n_nh^3}\right)\ ,
\end{equation}

 \noindent where $\mu_e$=$(3\pi^2Y_e\,\rho)^{1/3}$ is the chemical potential of the fully degenerate relativistic electrons, $m_N$=939.1 MeV is the neutron mass and $g_0$=2 (spins) is the statistical weight of the neutron.

\begin{figure}[h!]
\includegraphics[scale=0.5]{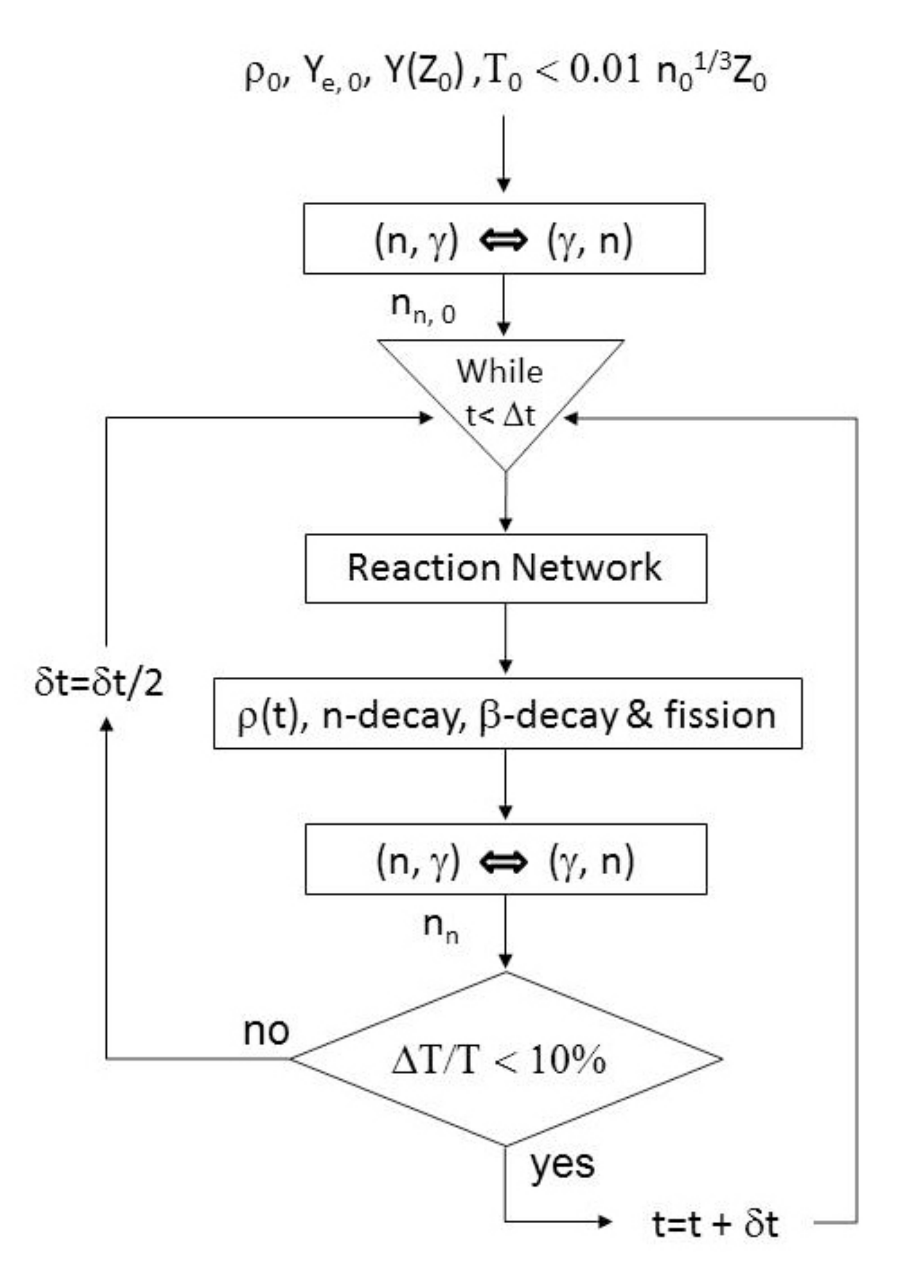}
\caption{Schematic representation of a single time step in our dynamic code.
In this chart $\rho_0$ is the initial crust density, Y$_{e,0}$ is the initial electron to baryon ratio, Y(Z$_0$) is the initial chemical abundance of the crust, T$_0$ is the initial temperature, Z$_0$ is the initial atomic number and n$_0$ is the initial particle density of the crust.  The neutron density (n$_n$) is first found using the secant method and then reaction network is solved using the Crank-Nicholson method.  After solving the reaction network evolution of the density ($\rho(t)$), heating due to $\beta$ and neutron decay and nuclear fission is calculated.  The neutron density is then re-evaluated and finally the maximal temperature change criterion is checked before moving on to the next time step.}
\label{Chart}

\end{figure}

 Once the code begins to run, at each and  every time step, we update the abundance of nucleus $i$ 
if the fractional change $dY_i/Y_i <  0.1$,  else we  decrease  the time  step  until  this condition  is satisfied.  From   Eqn.~(\ref{general_equation}),  this  implies  that $|\partial   n_i/n_i-\partial\rho/\rho|<0.1$.   For   non-relativistic expansion, both $\partial  n_i/n_i$ and $\partial\rho/\rho$  are small, but for  relativistic expansion,  $\partial\rho/\rho$ can be  large at early times. We can still ensure  that the condition on the $Y_i$'s is
satisfied    by   choosing    our   time    step    small   enough that
$\partial\rho/\rho< 0.1$,  provided $\partial\rho/\rho$ dominates over
$\partial  n_i/n_i$. This  is  equivalent  to an  upper  limit on  the
entropy  which  can   be  converted to an upper limit on the initial temperature $T_{\rm up}$(K)=0.01$n_0^{1/3}Z_0$, where $n_0$ is the initial ejecta number density (cm$^{-3}$). For example, taking $Z_0=Z_{Fe}$ (iron-rich ejecta), we find that an initial density $\rho_0\sim 10^{13}$g/cc requires that the initial temperature satisfy $T_0 \lesssim 10^{11}$K.  If the temperature input by the user violates this density relation r-Java asks for a new choice of initial temperature and shows the user the upper bound.  Introducing heat  into the
system  increases   the  entropy  and  can potentially reduce   the  maximum  initial temperature    for   which    our   abundance    criterion    can   be
satisfied.  However,  heating due  to  nuclear  transmutations can  be
ignored at early  times in the case of fast relativistic  expansion, since the
expansion is much faster than typical beta-decay time-scales. Thus, the
above estimate for  the maximum initial temperature remains  valid for fast
expansions.  It is  automatically satisfied  for  the non-relativistic
case,  where $\partial\rho/\rho$  is much  smaller.  Furthermore, once
satisfied  at the  outset, the  abundance criterion  $dY_i/Y_i<0.1$ is
seen to be always satisfied for all future times. These features are naturally
built-in to our code, as detailed by the algorithm flow-chart seen in Figure \ref{Chart}.  

Equation (\ref{reacnet}) for the evolution of Y(Z) must now be modified due to the density evolution, which was neglected before in the static calculations. The total derivative of $Y(Z)$ is

\begin{equation}
\label{general_equation}
 \frac {d Y_{i}}{dt} = n_{i} \left(\frac{\partial{\frac{1}{\rho}}}{\partial{t}}\right)_{n_{i}={\rm const}} + \frac{1}{\rho } \left(\frac{\partial{n_{i}}}{\partial{t}}\right)_{\rho = {\rm const}} \ .
\end{equation}

\noindent The first term, which describes changes in ejecta density, can be rewritten in term of $Y_{i}$ and then incorporated into our reaction network:

\begin{equation}
\label{yofz}
\frac{n_{i}}{N_{A}} \frac{\partial{\frac{1}{\rho}}}{\partial{t}} = -\frac{n_{i}}{N_{A}} \frac{1}{\rho ^2} \frac{\partial{\rho}}{\partial{t}} = - \frac{1}{\rho} \frac{\partial{\rho}}{\partial{t}} Y_{i} \ .
\end{equation}

\noindent For the dynamic simulation, we used a Crank-Nicholson method to solve Eqn.(\ref{yofz}). As in the static case, we were able to solve the linear system of abundance equations to find the variations in abundances within a timestep. These nuclear transmutations generate entropy and lead to heating, so to keep track of this, we inverted the entropy from these changes to compute the change in temperature. This procedure was implemented in a self-consistent manner, i.e, the new temperature was fed back to check the value of entropy. The heating of the material via $\beta$-decay is added to Eqn.(\ref{Entropy}) as $\delta s = \delta q / T $ and this whole expression is inverted to get the new temperature after each time step. We check that the temperature remains above 2$\times 10^9$K at all times during r-process, so that the waiting point approximation we employ is still valid.

\begin{figure*}[t!]
\includegraphics[width=\textwidth]{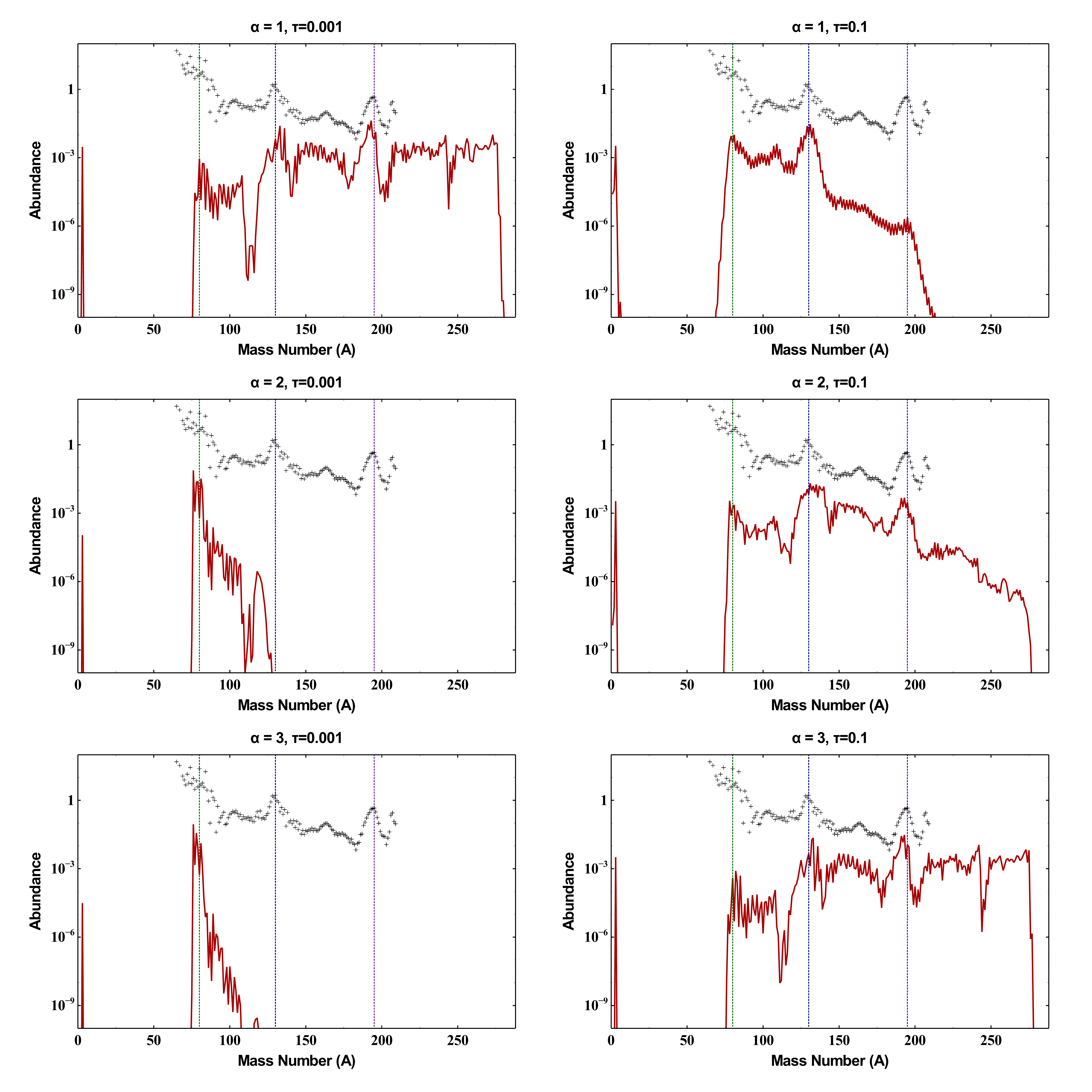}
\caption{The final abundances (solid red line) for dynamic simulations (one second duration) of nucleosynthesis from expansion of a chunk of pure Iron situated in neutron-rich matter with $Y_e=0.03$. Solar abundances are shown as black crosses.  For each panel the same initial temperature ($4\times10^9$K) and density ($1\times10^{11}$g/cc) is used.  The left column of panels use an expansion time-scale of $\tau=0.1$s and for the right column of panels the expansion time-scale is $\tau=0.001$s.  Each row considers a different density profile: {\bf top} $\rho(t)\propto \left(t/\tau\right)^{-1}$, {\bf middle} $\rho(t)\propto \left(t/\tau\right)^{-2}$, {\bf bottom} $\rho(t)\propto \left(t/\tau\right)^{-3}$.}
\label{dynsim}
\end{figure*}

\begin{figure*}
\includegraphics[width=\textwidth]{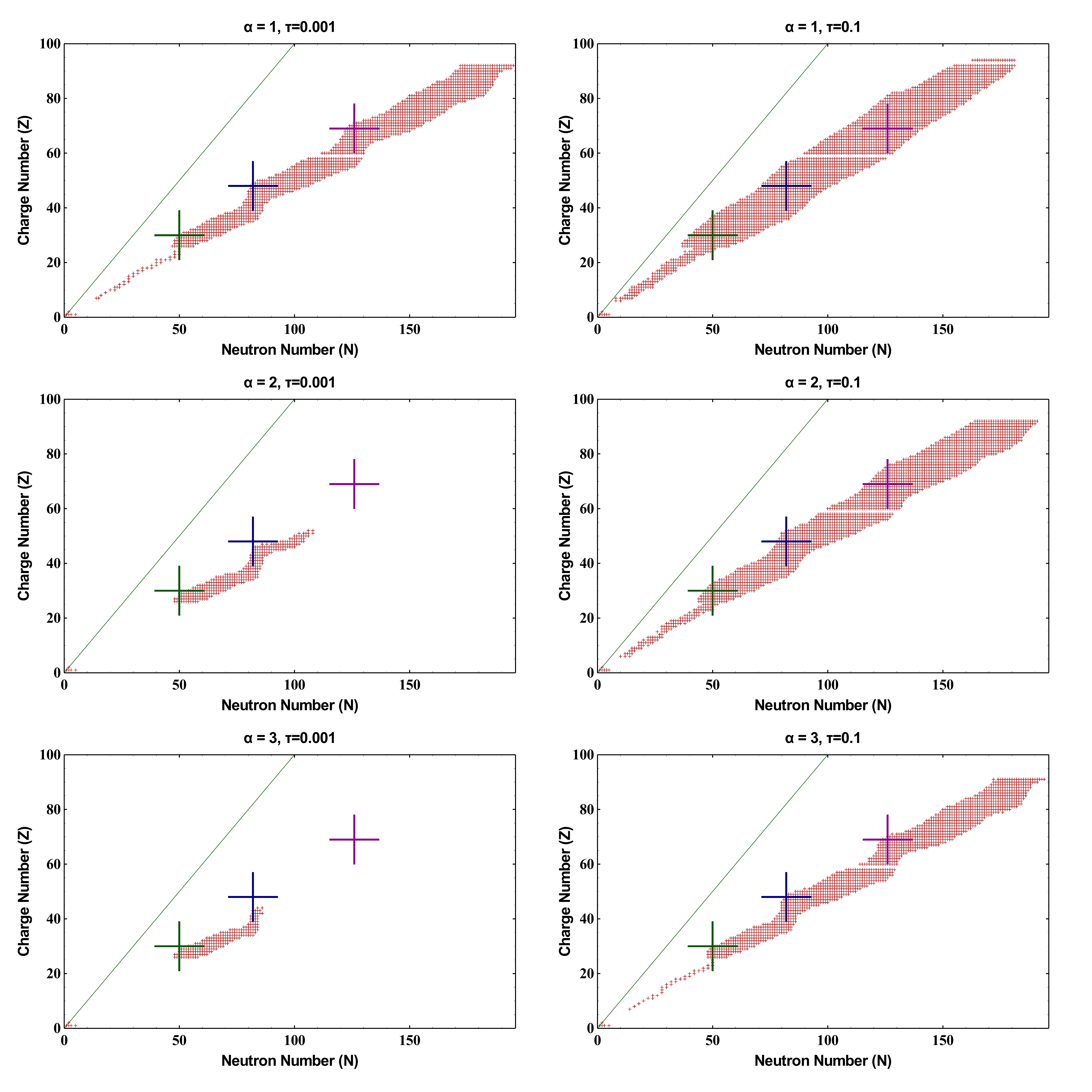}
\caption{The final Z versus N distribution is plotted for each dynamic simulation (after one second duration).  In each panel the solid green line is at $N=Z$, and the large crosses mark the location of the closed neutron shells.  For each panel the same initial temperature ($4\times10^9$K), density ($1\times10^{11}$g/cc) and electron fraction (0.03) are used.  The left column of panels use an expansion time-scale of $\tau=0.1$s and for the right column of panels, the expansion time-scale is $\tau=0.001$s.  Each row considers a different density profile: {\bf top} $\rho(t)\propto \left(t/\tau\right)^{-1}$, {\bf middle} $\rho(t)\propto \left(t/\tau\right)^{-2}$, {\bf bottom} $\rho(t)\propto \left(t/\tau\right)^{-3}$. }
\label{dynZN}
\end{figure*}

\begin{figure*}
\includegraphics[width=\textwidth]{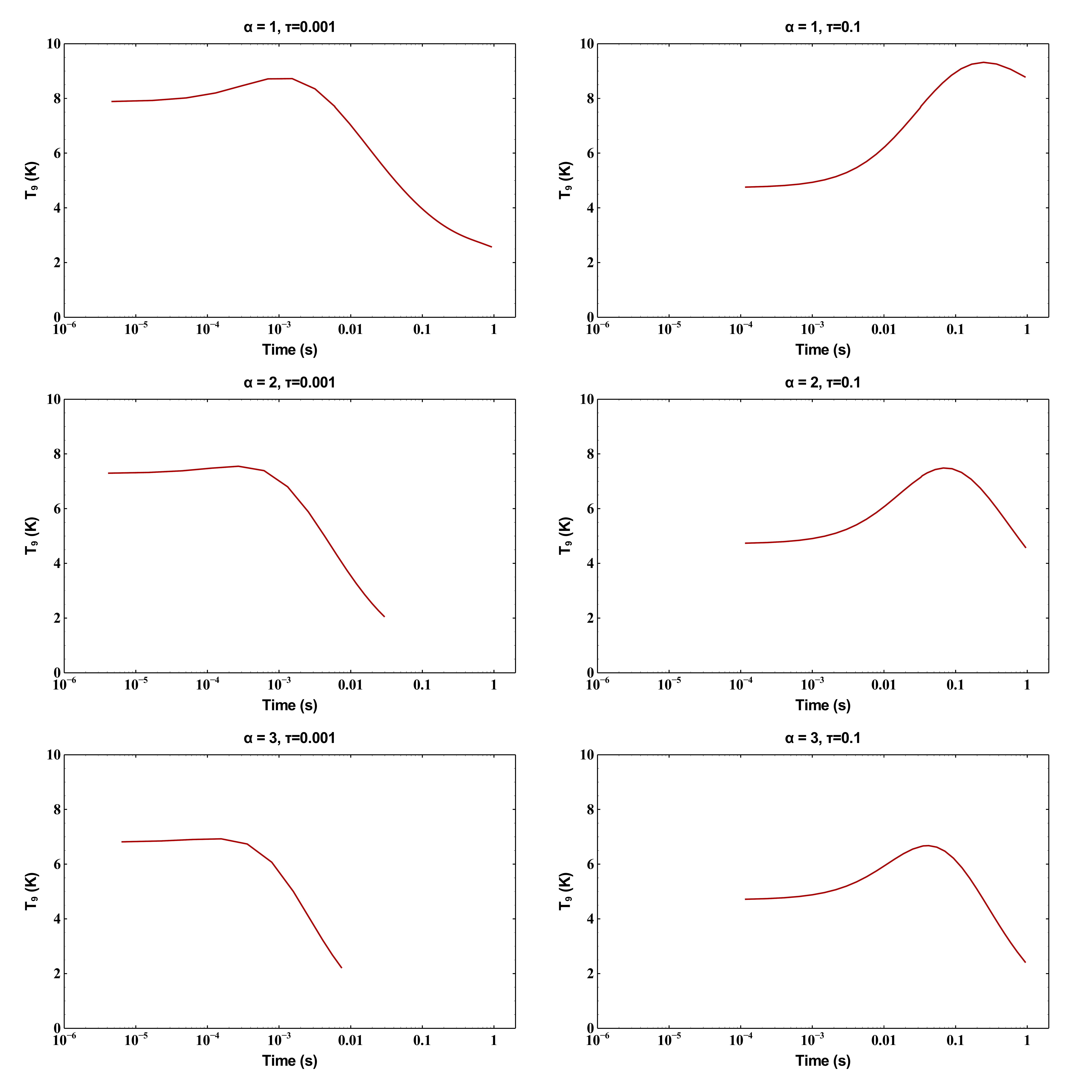}
\caption{The evolution of temperature over the simulation run is plotted for each dynamic simulation (50\% heating from $\beta$-decays is assumed).  For each panel the same initial temperature ($4\times10^9$K), density ($1\times10^{11}$g/cc) and electron fraction (0.03) are used.  The left column of panels use a expansion time-scale of $\tau=0.1$s and for the right column of panels the expansion time-scale is $\tau=0.001$s.  Each row considers a different density profile: {\bf top} $\rho(t)\propto \left(t/\tau\right)^{-1}$, {\bf middle} $\rho(t)\propto \left(t/\tau\right)^{-2}$, {\bf bottom} $\rho(t)\propto \left(t/\tau\right)^{-3}$.  }
\label{dyntemp}
\end{figure*}

To complete the time step, we also need to explicitly calculate the neutron density, which we can do using $(n,\gamma)\rightleftharpoons(\gamma,n)$ equilibrium, as long as temperatures and densities remain high enough. This ensures equilibrium within any isotopic chain, which will instantly redistribute $Y(Z)$ between all isotopes in accordance with the neutron separation energy. Once again, we use the population coefficients which depend on the neutron density, so to get the equilibrium value of neutron density at each timestep, we solve the implicit equation of mass conservation:

\begin{equation}
\label{nn_equation}
 n_{n} m_{n} = \rho N_{a} \left(1 - \sum A(n_{n})\right) \ ,
\end{equation}

\noindent where $\sum A$ is the mass of all nuclei {\it except} neutrons. This quantity depends on the neutron density through the population coefficients, which give the effective mass of every isotopic chain. We used a secant method to solve this equation and get the equilibrium value of $n_{n}$. Then the system is completely described and the code can move to the next time step. We can run the code as long as the waiting point approximation is valid, for $n_{n} > 10^{20} cm^{-3}$ and $T > 2\times 10^{9} K$ (see algorithm in Figure \ref{Chart}).

\subsection{Dynamic Simulation Results}

In Fig. \ref{dynsim}, we explore the effect on final abundance of changing the density profile of the expanding material.  To do this we considered the following general density profile:

\begin{equation}
\rho \left(t \right) = \frac{\rho_0}{\left(1+\frac{t}{\tau}\right)^{\alpha}} \, .
\end{equation}

\noindent For Fig. \ref{dynsim}, Fig. \ref{dynZN} and Fig. \ref{dyntemp} in the top panels $\alpha=1$, the middle panels $\alpha=2$ and the bottom panels $\alpha=3$, where increasing values of $\alpha$ roughly simulate increasing degrees of freedom for expansion.  The left column of these figures use an expansion time-scale of $\tau=0.001$s while for the right column $\tau=0.1$s.  We left the remaining input parameters the same for all panels ($T_0=4\times10^9$K, Y$_{e,0}$=0.03, Z$_0$=26 and a simulation duration of one second).  From the left column of Fig. \ref{dynsim} it can be seen that the density drops off too rapidly to produce elements heavier than A=120 in all but the $\alpha=1$ case.  This is because the expansion is too fast for neutron captures to occur for a sufficiently long period of time to make heavy elements.  An opposite trend can be seen in the $\tau=0.1$s cases where the heavy element production increases with the degrees of freedom for expansion.  This is because for slow expansions, the density, particularly for low degrees of  freedom ($\alpha$=1) remains high for an extended period of time, blocking beta-decays that are needed to move the flow up to higher mass numbers.
Consequently, the third peak is absent or weak. For $\alpha$=2 or 3, the density fall off is rapid enough to compensate this effect, so a strong 3rd peak is seen.  The effect of the beta-decay blocking can be clearly seen in the Z vs. N plots shown in Fig. \ref{dynZN}.  In the top panel, right column of Fig. \ref{dynZN}, it is clear that the slower expansion case allows elements to build up beyond the magic numbers, while for faster expansion (top-left panel of Fig. \ref{dynZN}) or multiple degrees of freedom (right column of 2nd and 3rd panel) there exists clear steps at each magic number, revealing the imprint of the extra stability. It is interesting to note that the final abundances seen in the top left ($\tau=0.001$, $\alpha=1$) and bottom right ($\tau=0.1$, $\alpha=3$) panels of both Fig. \ref{dynsim} and \ref{dynZN} are remarkably similar.  This similarity is due to the fact that after one second of expansion the effects of the different $\alpha$ and $\tau$ values will directly offset one another leaving the final density of the expanding chunk the same.  The slight difference in final abundances of the two simulations are due to temperature effects.  Finally, Fig. \ref{dyntemp} displays the temperature evolution for each of our dynamical simulations.  The left column of Fig. \ref{dyntemp} exemplifies how in the short expansion time-scale adiabatic cooling dominates the heating from beta-decays, while in the longer expansion time-scale cases (the right column of Fig. \ref{dyntemp}) beta decays have a strong heating effect early on, specially for $\alpha$=2,3. 

\section{Discussion and Conclusion}

We have constructed and tested a nucleosynthesis code for the r-process under static and dynamic conditions. Called r-Java, it is made available online through an easy-to-use interface at http://quarknova.ucalgary.ca/ . The user is given a number of input options as well as the ability to easily modify the nuclear data used in the code. In this paper, we have studied NSE, as well as both static and dynamic simulations.

Exploring static simulations for a range of neutron irradiation and temperatures, we found that neutron density $\sim 10^{28}$cm$^{-3}$ and above is required for a strong third r-process peak. With increasing temperature beyond $T_9$=2, nuclei photo-disintegrate extremely fast, before significant pile-up can occur at the true waiting-point nuclei. This resulted in abundance peaks at smaller mass numbers. This feature is compensated by increasing the neutron density; however, the effect of fission will then play a role in determining the final abundances. We intend to include fission-rates in our static simulations in the future, so that higher neutron densities can be reliably addressed. We have included a $\beta$-decay module that allows the user to track the flow of $\beta$-decays once neutron bombardment stops. This should be useful in studying freeze-out conditions and quasi-equilibrium reaction networks.

Using the dynamic module of r-Java we compared different density profiles for two expansion time-scales.  We found that at the shorter expansion time-scale ($\tau=0.001$s) only the slowest evolving density profile ($\rho(t)\propto \rho_0\left(t/\tau\right)^{-1}$) produced a significant abundance of elements above A=120.  For the longer expansion time-scale ($\tau=0.1$s) the heavy abundance yield increased with faster evolving density profiles, where the strongest A=195 peak occurred in the simulation that considered $\rho(t)\propto \rho_0\left(t/\tau\right)^{-3}$.  At present, all our calculations are carried out in the waiting-point approximation.  

The large qualitative differences in the r-process pattern we observe as a function of dynamical expansion parameters are important for alternate sites of the r-process, such as neutron star mergers or spherically symmetric decompressing neutron star matter. In the case of mergers, tidal forces can lead to a "tube of toothpaste" quasi-1D ejection, whereas spherical decompressions resemble the 3D case. From Fig.\ref{dynsim}, we see that fast expansion is required if heavy elements are made in mergers, while less energy (slower expansion) is required for spherical decompression. This is the case, for eg., in Quark-Novae, which
will be discussed in a follow-up paper. We also note the curious fact that the peak at A = 3, corresponding to Tritium, is formed from the sequence of reactions comprised of neutron decay followed by formation of Deuterium and then Tritium. Since the half-life of Tritium is relatively short (12.32 years), a large amount of its daughter element, He-3, would be formed and dispelled into the surrounding space. In the case of a dual-shock Quark-Nova (\cite{LO,ouyed2010}), this can have important consequences, through spallation on supernova ejecta, for cosmic rays and anomalous abundance ratios in some supernovae (eg., Cas A).

Our future efforts are aimed in three directions: (1) applications of r-Java to Neutron Star Mergers and the Quark-Nova scenario. The Quark-Nova is an ejection mechanism for neutron star material powered by the phase conversion from hadronic to deconfined quark matter occurring at the core of a neutron star~(\cite{ODD,KOJ}). The extreme energetics and high neutron-to-seed ratio makes the Quark-Nova an ideal candidate for an astrophysical r-process site~(\cite{Jai07}); (2) including the effects of freeze-out through a set of quasi-equilibrium reaction networks and other processes neglected in Eq.(\ref{eq:Yevolv1}) such as beta-delayed neutron emission, to obtain a more realistic pattern of final abundances, which can then be used for schematic chemical evolution studies; (3) based on~\cite{Nieb}, further numerical study of the dynamics of the nuclear-quark conversion front inside the neutron star to better constrain the Lorentz factor of the ejecta, and possibly develop a semi-analytic approach to constrain the parameter space, similar to what has been done for neutrino-driven winds from supernovae.

\section{Acknowledgments}  RO, MK and NK are supported by the Natural Sciences and Engineering Research Council of Canada.  NK acknowledges support from Alberta Ingenuity and the Killam Trusts. PJ acknowledges start-up funding from California State University Long Beach. 


\end{document}